\documentclass[11pt]{elsarticle}

\usepackage{amssymb}

\usepackage{xcolor}
\usepackage{graphicx}
\usepackage{float}
\usepackage[margin=2.5cm]{geometry}

\graphicspath{ {./Figures/} }

\begin{document}

\begin{frontmatter}



\title{From seeing to remembering: Images with harder-to-reconstruct representations leave stronger memory traces}


\author[inst1,inst4, inst5,inst7]{Qi Lin}
\author[inst2,inst5]{Zifan Li}
\author[inst2,inst3,inst6,inst7]{John Lafferty}
\author[inst1,inst2,inst3,inst6,inst7]{Ilker Yildirim}

\affiliation[inst1]{organization={Department of Psychology},
            addressline={Yale University}}
\affiliation[inst2]{organization={Department of Statistics and Data Science},
            addressline={Yale University}}
\affiliation[inst3]{organization={Wu Tsai Institute},
            addressline={Yale University}}
\affiliation[inst4]{organization={Center for Brain Science}, addressline={RIKEN}}
\affiliation[inst5]{These authors contributed equally.}
\affiliation[inst6]{These authors jointly supervised this work.}
\affiliation[inst7]{Correspondence should be addressed to these authors.}

\begin{abstract}

Much of what we remember is not due to intentional selection, but simply a by-product of perceiving. This raises a foundational question about the architecture of the mind: How does perception interface with and influence memory? Here, inspired by a classic proposal relating perceptual processing to memory durability, the level-of-processing theory, we present a sparse coding model for compressing feature embeddings of images, and show that the reconstruction residuals from this model predict how well images are encoded into memory. In an open memorability dataset of scene images, we show that reconstruction error not only explains memory accuracy but also response latencies during retrieval, subsuming, in the latter case, all of the variance explained by powerful vision-only models. We also confirm a prediction of this account with `model-driven psychophysics'. This work establishes reconstruction error as a novel signal interfacing perception and memory, possibly through adaptive modulation of perceptual processing. 

\end{abstract}
\end{frontmatter}

\section*{Introduction}
So much of what we remember is not the result of intentional selection, but rather the result of simply perceiving. How are perceptual experiences cast into memory? And how does perceiving exert control over remembering? These are fundamental questions in the study of the mind with multiple lines of empirical and theoretical studies designed to uncover the interface between perception and memory (e.g., \cite{wagner1998building, xue2018neural, craik1972levels,schurgin2020psychophysical,chun2011memory,kurby2008segmentation,favila2020transforming,liu2021transformative,libby2021rotational, serences2016neural,xu2017reevaluating}). A striking illustration of the extent to which perception influences memory is the recent demonstration of `memorability', the finding that some images are systematically more memorable than others  across observers \cite{isola2014makes, bainbridge2013intrinsic}. Formation of new visual memory traces must recruit both visual and memory-related functions, but the computational basis of how they interact to produce memory traces remains poorly understood. 

Existing computational accounts, inspired by the demonstration of image memorability, largely consider models that involve vision-only computations, such as the deep convolutional neural networks (DCNN) trained for image classification \cite{jaegle2019population, khosla2015understanding, lin2019image, kramer2022features}. These studies have established a quantitative relationship between the summary statistics derived from the later stages of these networks (e.g., the magnitude of activations of a given layer) and memorability scores of images. Interestingly, this effect is also observed neurally: the population response magnitude of the inferior temporal cortex neurons tracks the memorability scores of the presented images. However, these vision-only models do not attempt to formalize processes responsible for transforming percepts into memories and thus remain incomplete as computational accounts of how perceptual processing relates to memory traces. 

A classic psychological account, the `level-of-processing' theory of Craik and Lockart \cite{craik1972levels}, has attempted to directly  address this interface --- 
proposing that a memory trace is a by-product of the perceptual analysis of the incoming sensory signals and that a `deeper' analysis is associated with better retention in memory. 
However, this framework, including a specification of what determines the depth (or level) of perceptual processing \cite{baddeley1978trouble,treisman2014psychological,craik2020remembering}, remains largely qualitative and non-computational. 
To date, all empirical demonstrations of the level-of-processing effect rely on the use of an orientating task (as reviewed in \cite{cermak2014levels}), completely missing the automatic nature of perceptual processing. In fact, work investigating how orientating tasks interact with image memorability has demonstrated that the effect of such orienting tasks is independent of memorability \cite{bainbridge2020resiliency}. 
A computational form of the level-of-processing theory must take into account perceptual processing more rigorously and address how its depth can be modulated on an image-by-image basis. 

By addressing the elemental computations thought to be underlying memory -- compression \cite{bates2020efficient} and reconstruction \cite{schacter2012adaptive, hemmer2009bayesian} --, we present a new computational model that yields a stimulus-driven, quantitative measure for how perceptual processing can impact memory formation. 
This model combines the vision-only models mentioned above with a sparse coding framework (SPC), a classic architecture used for compressing information in both computational neuroscience (e.g., \cite{olshausen1996emergence,olshausen1997sparse, benna2021place,lewicki2002efficient}) and machine learning (e.g., \cite{zemel1993developing,rozell2008sparse,rumelhart1985learning,gregor2010learning}). Our model operates on the activations evoked by natural scene images in a DCNN trained to categorize scenes and objects \cite{zhou2017places, DBLP:journals/corr/SimonyanZ14a} and learns how to reconstruct these evoked activations. When considered over the entire space of signals to be compressed, reconstruction error, measured as the difference between the signals recovered from compressed codes and the uncompressed signal, provides a benchmark for evaluating different codes for lossy compression \cite{berger1971rate, cover1991elements, mackay2003information,hinton1995wake}. 

We hypothesize that reconstruction error provides the necessary computational substrate for gauging and modulating the level of perceptual analysis \cite{craik1972levels}, and thus impacts memory strengths. In particular, the reconstruction error resulting from the sparse coding model provides a principled signal to determine how much more processing might be warranted on an image-by-image basis. Ideally, a valid computational account addressing the perception-to-memory interface should capture different aspects of memory behavior, targeting not just the accuracy of retrieval, but also systematic variance in terms of latencies during retrieval (i.e., response time), in addition to predicting measurable new phenomena. 

Across three studies, this work aspires to this ideal by testing our sparse coding model's ability to capture novel aspects of how sensory signals are transformed into memory, above and beyond what can already be explained by vision-only models. In Studies 1 and 2, we focus on two well-established and complimentary measures of memory strength: memory accuracy and response times during retrieval \cite{kahana1999response}. To this end,  we relate the reconstruction errors of images (obtained using the sparse coding model) to their memorability scores and response times measured in a large scene memorability dataset\cite{isola2014makes} while taking into account what can be explained by a standard DCNN trained for image classification (VGG-16; \cite{zhou2017places}). We find that reconstruction error explains additional variance in both memory accuracy (Study 1) and response times (Study 2), subsuming all of the variance explained by other models in the latter case. In Study 3, we then turn to `model-driven psychophysics' and predict that the architectural differences between the sparse coding model and DCNNs would be paralleled in the brain as temporally and functionally distinct processes. 
In a pre-registered experiment, we manipulate the encoding times in a rapid serial visual presentation (RSVP) paradigm and observe that images with large reconstruction error benefit more from longer encoding times, while controlling for DCNN-driven memorability effects.
Together, these results establish compression-based reconstruction error as a previously unrecognized driver of memorability, and suggest a mechanism in which such reconstruction error modulates the depth of encoding of the incoming visual inputs.

\section*{Results}
\label{sec:results}

\begin{figure}[hbt!]
  \centering
  \includegraphics[width=1\textwidth]{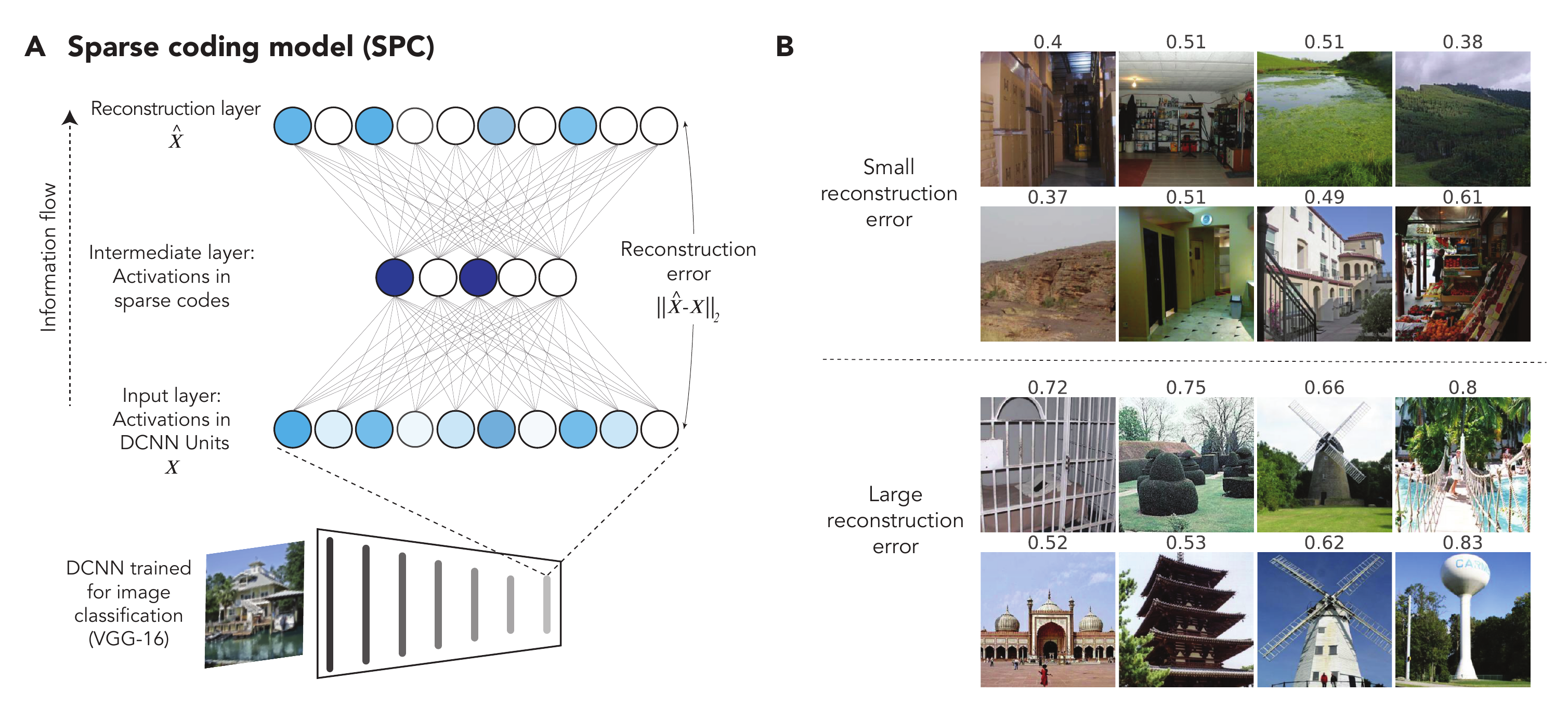}
  \caption{Schematic representation of how reconstruction errors are quantified using a sparse coding model (SPC). The sparse coding model example shown here is based on Layer 7 of a deep convolutional neural network (DCNN) trained for image classification --- the second fully-connected layer from the VGG-16 network; \cite{zhou2017places}. (B) Example images that are easy and hard for the SPC to reconstruct. The numbers above the original images are the memorability scores (hit rates - false alarm rates) as measured in \cite{isola2014makes}. } 
  \label{fig:fig1}
\end{figure}

\subsection*{Efficient compression of sensory activations by minimizing reconstruction error}
To model the process of compressing sensory activations underlying our visual experiences, we built a sparse coding model (in short, SPC)
to reconstruct the activations in a given layer of a DCNN pretrained to classify images of scenes and objects\cite{zhou2017places} based on the commonly used VGG-16 architecture\cite{DBLP:journals/corr/SimonyanZ14a}. 
The SPC model (see Fig. 1A for a schematic) consists of three layers: an input layer of 1000 units, an intermediate layer of 500 units for recoding the inputs, and a reconstruction layer of 1000 units for reproducing the input. 
We sampled 7 layers (ranging from early to late) from the VGG architecture and trained a separate sparse coding model for each of these DCNN layers.
Because the dimensionality of DCNN layers is typically very high, we designed the input layer of the SPC (and thus, the reconstruction layer) to consist of 1000 units randomly sampled from the corresponding DCNN layer. This design also ensured that the number of parameters to be trained in the SPC model remains constant despite the fact that the dimensionality of layers varies across the DCNN architecture. 
Our SPC model is trained to reconstruct the DCNN activations evoked by natural scene images (10263 images in total) from a publicly available dataset \cite{isola2014makes}, allowing the model to learn an efficient code for reconstructing DCNN activations for complex visual inputs. 
Critically, unlike the training of DCNN where the goal is to maximize classification accuracy, the objective of SPC training is to minimize the reconstruction error (the Euclidean distance between the reproduced and original activations, subject to a sparsity term on the intermediate recoding layer; see Fig. 1A and Methods).    

\subsection*{Study 1: Images with larger reconstruction error are more memorable}

After training, sparse coding yields substantial variability in how well individual images can be reconstructed (see Fig. S1). In Fig. 1B, we show example images that are easy and hard for a representative SPC model (trained on the Layer 7 activations in the DCNN) to reconstruct. 
We observe that the hard-to-reconstruct images tend to be more object-centered or contain humans, relative to the easy-to-reconstruct images. Interestingly, such image attributes have been related to memorability behavior in earlier work \cite{isola2014makes, bylinskii2015intrinsic}. In fact, even from this subset of images selected solely based on their reconstruction error, we observe that hard-to-reconstruct images are often more memorable (memorability scores\footnote{The memorability score of a given image is calculated as hit rate minus false alarm rate measured in \cite{isola2014makes}. Hit rate is calculated as the proportion of observers who correctly indicated that the image was repeated when the image was shown a second time to these participants. False alarm rate is calculated as the proportion of observers who incorrectly indicated that the image was repeated when the image was shown for the first time to these observers. In other words, a higher score means an image is more memorable after accounting for general familiarity.} indicated above each image). Based on these observations, we hypothesize that the images that are harder to reconstruct are also more memorable. 

To establish this prediction quantitatively, we focused on the 2221 target images with available memorability scores measured in \cite{isola2014makes}.  
For each of the 2221 target images, we obtained a reconstruction error from each of the 7 trained SPC models. Then for each SPC model, we correlated the resulting reconstruction errors and memorability scores of the corresponding images. As shown in Fig. 2A (left panel), reconstruction errors from all sampled layers in SPC were significantly related to memorability: images with larger reconstruction error are more memorable. Layers 5 (conv5) and 7 (fc2) showed the strongest effects. 

Since the SPC models were trained based on the activations from a feedforward VGG-16 DCNN pretrained to classify scenes and objects \cite{zhou2017places}, we next wanted to understand the extent to which the resulting reconstruction errors are just capturing the same variance in memorability as previous measures derived from purely feedforward architectures (e.g., \cite{jaegle2019population, lin2019image, kramer2022features}). To this end, we derived a predictor from the DCNN network by sampling the same 7 layers as the ones used for training the SPC models. Following \cite{lin2019image}, for each layer and target image, we calculated the distinctiveness as the Euclidean distance between each target image and its nearest neighbor with respect to the DCNN's feature space at this layer. Replicating similar results \cite{jaegle2019population}, we found that distinctiveness in the DCNN across all sampled layers was significantly related to memorability, with the later layers (Layers 5-7) showing the strongest effects (see Fig. 2B, left panel). 

\begin{figure*}
  \centering
  \includegraphics[width=1\textwidth]{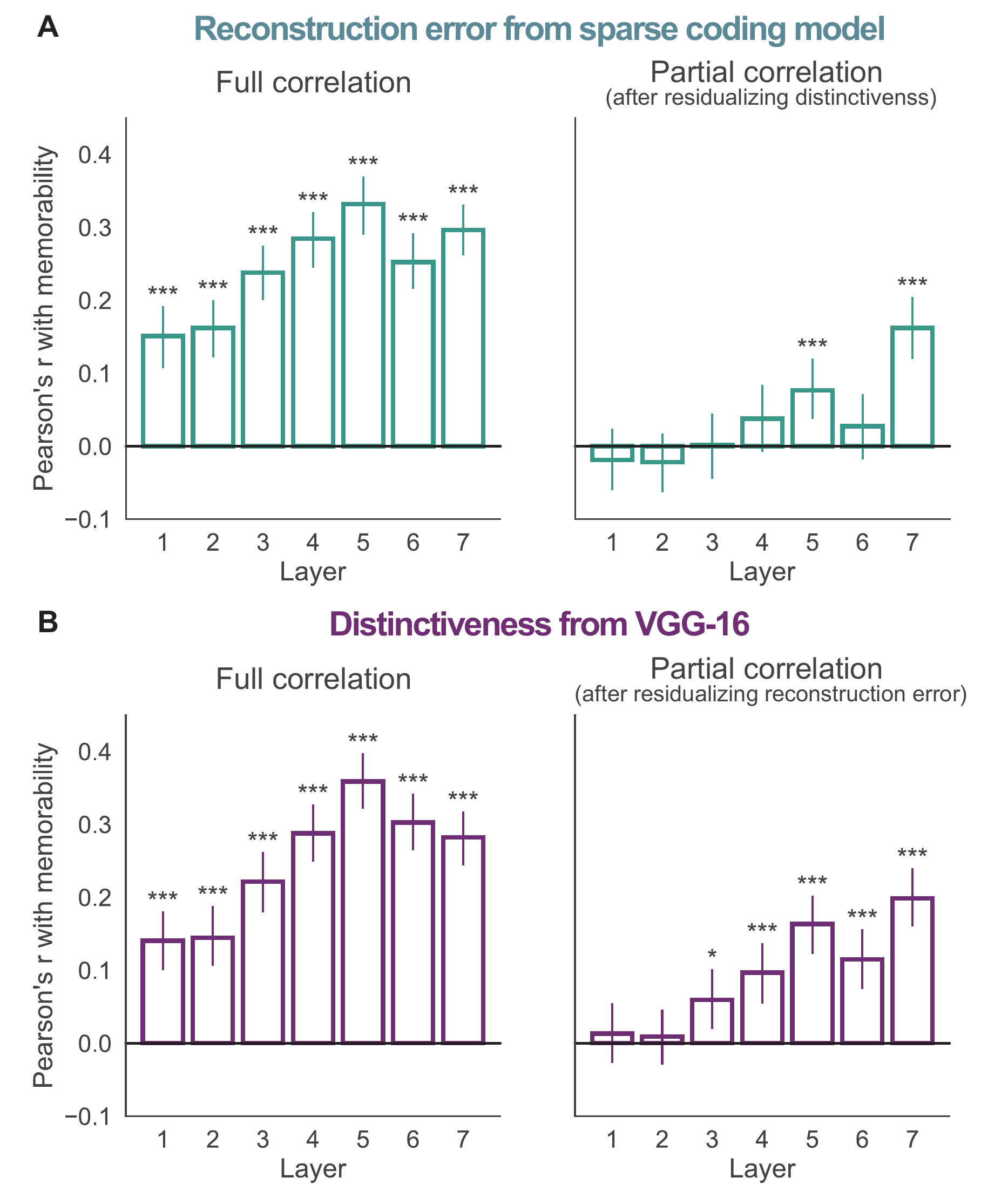}
  \caption{Images with large reconstruction error are more memorable. (A) Pearson's r between memorability and reconstruction error and partial Pearson's r after accounting for distinctiveness. (B) Pearson's r between memorability and distinctiveness and partial Pearson's r after accounting for reconstruction error. Error bars represent 95\% confidence intervals from 1000 bootstrapping iterations. ***: \(p <.001\), **: \(p <.01\), *: \(p <.05\). P values have been corrected for multiple comparisons with Bonferroni correction (\(\alpha=0.05/14\)).}
  \label{fig:fig2}
\end{figure*}

After establishing that both reconstruction error and distinctiveness are significantly correlated with memorability, we next asked: does reconstruction error capture additional variance in memorability, above and beyond what was already explainable by the DCNN's feature hierarchy optimized for image classification?  
To address this question, we chose distinctiveness from Layer 5 as our primary measure of distinctiveness since it showed the highest correlation with memorability (Fig. 2B, left panel). We then compared distinctiveness at that layer to reconstruction error as well as another standard measure also derived from bottom-up feature hierarchy that shows correlation with memorability: the L2-Norm of activations in a DCNN layer given an image \cite{jaegle2019population}\footnote{We found that distinctiveness was always better correlated with memorability scores and these correlations were more robust to the specific choice of layer, relative to L2-Norm (Fig. S2), further justifying our choice of distinctiveness instead of L2-Norm.}. First, we observed that although distinctiveness and reconstruction error were correlated (r = 0.82), this correlation was not perfectly co-linear and it was significantly less than the correlation between the two DCNN-derived measures, distinctiveness and L2-norm (r = .99; Williams' t test: p \(< .001\); see Fig. S3). These results suggest that the SPC model and the DCNN might capture different aspects of the computations underlying the memorability behavior.

Second, we performed partial regression analysis to directly test for the unique contribution of reconstruction error to explaining memorability. Specifically, we residualized Layer 5 distinctiveness (i.e., the DCNN layer that is most predictive of behavior under distinctiveness) from both memorability and reconstruction error. We found that indeed reconstruction error residuals continue to explain significant variance in memorability (Fig. 2A, right panel): reconstruction error from Layers 5 and 7 in SPC was still significantly correlated with memorability, after accounting for what can be explained solely by distinctiveness. 

Even though we are primarily interested in evaluating whether reconstruction error can explain additional variance, for completeness, we also ran a similar analysis by regressing out Layer 5 reconstruction error (which is the most predictive reconstruction error measure of memorability of all the layers) from both memorability and distinctiveness. As shown in Fig. 2B (right panel), distinctiveness from Layers 3-7 also remained predictive of memorability after controlling for reconstruction error, again suggesting that distinctiveness and reconstruction error are capturing separable aspects of the variance in memorability. 

Together, these results demonstrate that images with harder-to-reconstruct DCNN activations are more memorable and that reconstruction error makes additional contribution to image memorability, above and beyond distinctiveness. 

\subsection*{Study 2: Images with larger reconstruction error are recognized faster during retrieval}

Previous work that explicitly manipulates the depth of encoding with different orienting tasks has found that a deeper level of encoding is associated with faster reaction times during retrieval (e.g., \cite{vincent1996relations, ragland2003levels, bainbridge2020resiliency}).
Thus, if our results from Study 1 have to do with a mechanism in which reconstruction error modulates the depth of encoding, then we predict that harder-to-reconstruct images will be retrieved more quickly. 

To this end, we analyzed the response time data from the correct recognition trials in \cite{isola2014makes} (see Methods on details on data inclusion criteria).
Indeed, we found that reconstruction errors from all 7 layers in our sparse coding model were negatively correlated with response times during retrieval (Fig. 3A, left panel), such that harder-to-reconstruct images are faster to retrieve. We observed a similar pattern, albeit to a lesser degree and only for Layers 3-6, when we tested distinctiveness (Fig. 3B, left panel). To dissociate the contributions of reconstruction error and distinctiveness to explaining the variance in response times during retrieval, we again used partial regressions as in Study 1. We focused on model layers where we observed a significant correlation in the full correlation analysis and therefore did not perform partial correlation for Layers 2 and 7 distinctiveness. In the SPC model, the negative relationship between response times and reconstruction error on Layers 5-7 remained significant after we regressed out Layer 6 distinctiveness (the most predictive measure of response times in the DCNN) from both response times and reconstruction errors. On the contrary, distinctiveness decoupled from behavior (Layers 3-6) after we regressed out Layer 7 reconstruction errors (the most predictive measure of response times in the SPC model) from both response times and distinctiveness. 

These results demonstrate the specificity of the relationship between reconstruction error and response time during retrieval. That is, although both larger distinctiveness and larger reconstruction error predict higher recognition accuracy, only reconstruction error predicts retrieval efficiency. Beyond a mere measure of visual processing, this finding is consistent with our hypothesis that the magnitude of reconstruction error directly modulates the process, in particular the depth, of encoding percepts into memory. 

\begin{figure*}
  \centering
  \includegraphics[width=1\textwidth]{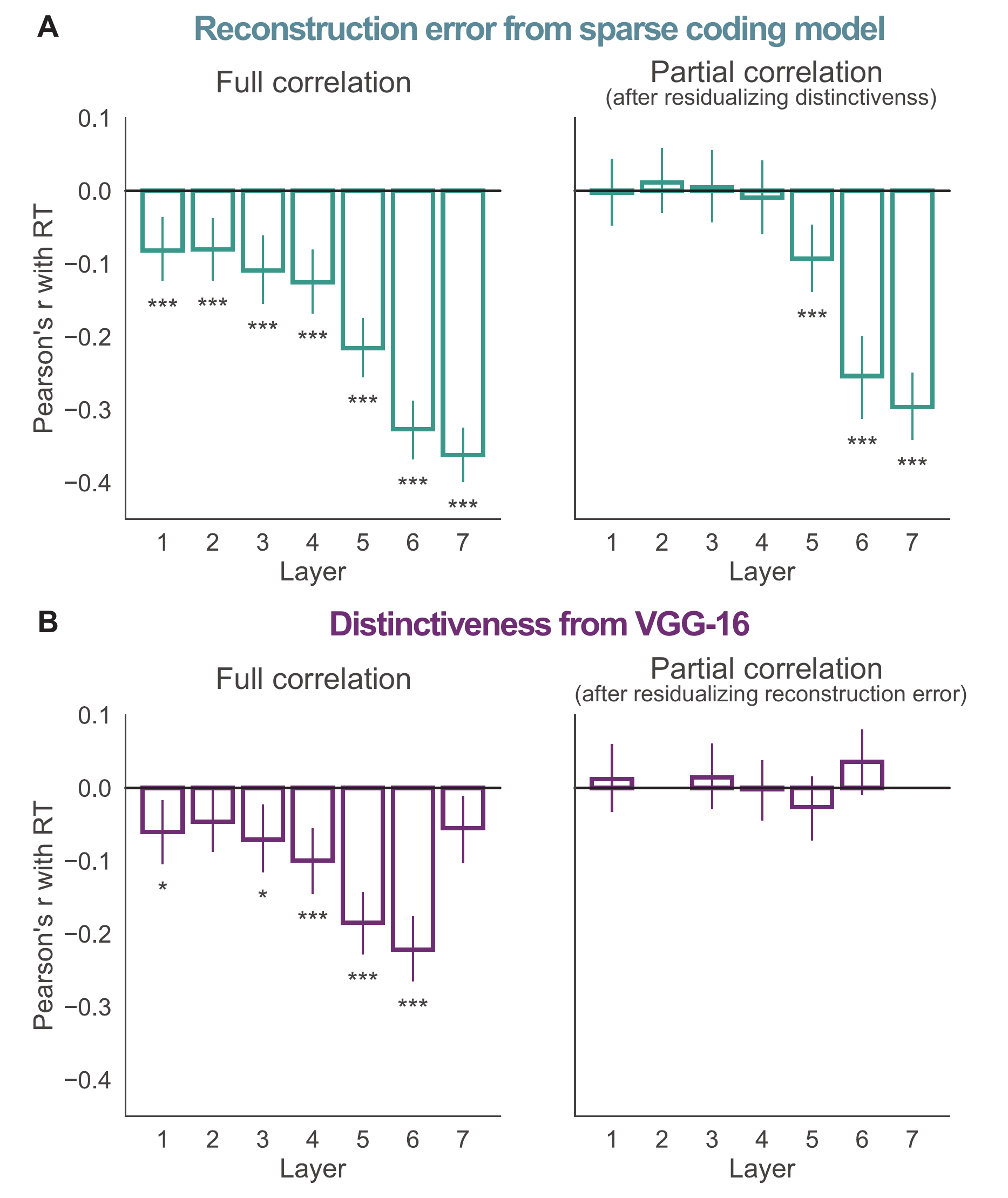}
  \caption{Images with large reconstruction error are recognized faster during retrieval. (A) Pearson's r between response times during retrieval and reconstruction error and partial Pearson's r after accounting for distinctiveness. (B) Pearson's r between response times during retrieval and distinctiveness and partial Pearson's r after accounting for reconstruction error. Note that partial correlation was only performed if the full correlation turned out to be statistically significant. Therefore, in the bottom right plot, there were no partial correlation results for Layers 2 and 7 distinctiveness.  Error bars represent 95\% confidence intervals from 1000 bootstrapping iterations. ***: \(p <.001\), **: \(p <.01\), *: \(p <.05\). P values have been corrected for multiple comparisons with Bonferroni correction (\(\alpha=0.05/14\)).}
  \label{fig:fig3}
\end{figure*}

\subsection*{Study 3: Images with larger reconstruction error benefit more from longer encoding times}

Studies 1 and 2 revealed reconstruction error as a novel source of image memorability and retrieval efficiency. These results suggest that reconstruction error might be modulating the process of encoding itself, in terms of how deeply an image should be encoded, and thus yielding these measurable relationships in a publicly available dataset. To further establish that reconstruction error is associated with a functionally distinct set of computations in the mind and brain, beyond the vision-only computations implemented in the non-linear feature hierarchies trained for image classification captured in distinctiveness, we turn to a prediction made by our modeling framework and test it in a `model-driven psychophysics' experiment in Study 3. 
Our logic is that images whose activations are harder to reconstruct and thus require a deeper level of processing will fetch additional mental resources,  manifested as longer encoding times and increased memory accuracy (\cite{craik1972levels}). 
Therefore,  we predict that the memory for images with higher reconstruction errors will benefit more from longer encoding times.  

To test this possibility, we start by sampling images with divergent profiles of distinctiveness and reconstruction error. Following Study 1 which focuses on memory accuracy, we used Layer 5 distinctiveness as our primary measure of distinctiveness and then selected Layer 7 reconstruction error from the SPC model as our primary measure of reconstruction error as this measure captures the largest amount of additional variance in memorability after accounting for distinctiveness (Partial Pearson's \(r\ = 0.16,\ p < .001\); Fig. 2A right panel)\footnote{To make sure that the Layer 7 reconstruction error is not just capturing the memorability driven by Layer 7 distinctiveness, we also tested including both Layer 5 and Layer 7 distinctiveness measures in the partial regression model. Even under this stringent way of controlling for distinctiveness, Layer 7 reconstruction error continued to significantly capture additional variance in memorability (Partial Pearson's \(r\ =0.08,\ p < .001\)).}. After settling on the primary measures of distinctiveness and reconstruction error, we then sampled four different groups of 48 images each: images with 1) large distinctiveness and large reconstruction error, 2) large distinctiveness and small reconstruction error, 3) small distinctiveness and large reconstruction error, and 4) small distinctiveness and small reconstruction error. `Large' is defined as falling within top 30 percentile of the target measure (distinctiveness or reconstruction error) and `small' is defined as bottom 30 percentile. See Fig. 4 for example images in each of the four groups.

\begin{figure*}[hbt!]
  \centering
  \includegraphics[width=1\textwidth]{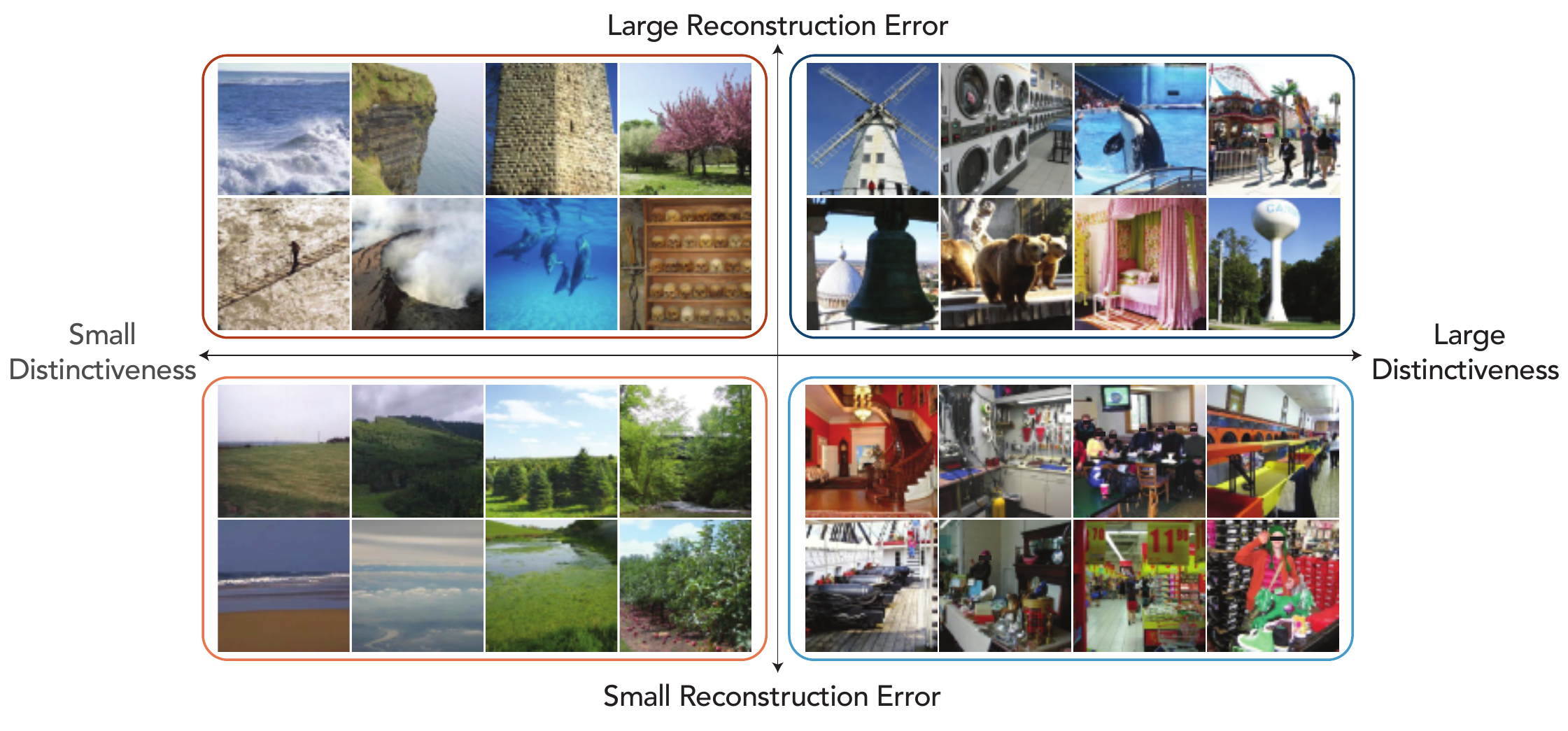}
  \caption{Example images from each of the four groups with different Distinctiveness-Reconstruction Error profiles.}
  \label{fig:fig4}
\end{figure*}

We adopted a within-participant design with 2 distinctiveness levels (Large vs. Small) × 2 reconstruction error levels (Large vs. Small) × 3 encoding durations (34, 84 or 167 ms). Each of the resulting 12 conditions was presented as a separate block for each participant. During each block, half of the trials were target-present trials (i.e., the test image was presented in the RSVP stream) whereas the other half were target-absent trials (i.e., the test image was not presented in the RSVP stream). On each trial, participants first saw an RSVP stream of images (see Fig. 5A; following the trial structure of \cite{broers2018enhanced}). Then they were shown a test image and had to indicate whether the test image was presented in the RVSP stream or not. The experiment design and sample size were pre-registered (https://aspredicted.org/MFM\textunderscore R22). 45 participants completed the experiment online via Prolific. We calculated hit rate \footnote{Target images were never used as foils in this RSVP experiment so we could not calculate false alarm rates for them.} for each of the 12 conditions (see Fig. 5B).

\begin{figure*}[hbt!]
  \centering
   \includegraphics[width=1\textwidth]{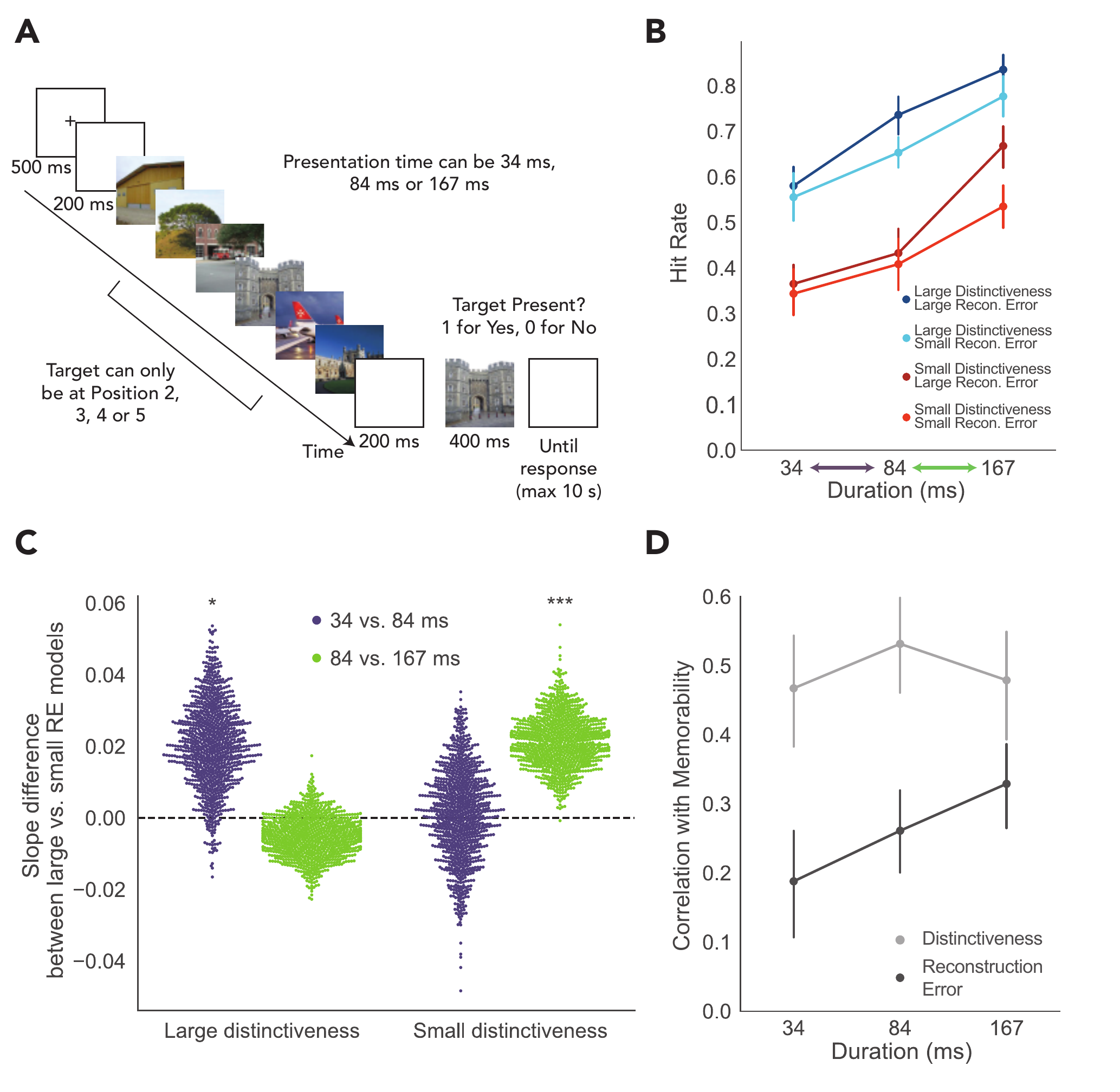}
  \caption{Images with harder-to-reconstruct representations benefit more from longer encoding times. (A) An example trial from the RSVP experiment. (B) Hit rates for each of the four image groups. Error bars represent 95\% confidence interval from 1000 bootstrapping iterations. (C) Difference between the regression slopes of hit rate vs. time across images with large vs. small reconstruction error, conducted separately for each sequential pair of encoding times (indicated by colors; also highlighted in panel B, x-axis) and distinctiveness level. Distributions represent the slope differences from 1000 bootstrapping iterations. ***: \(p <= .001\), *: \(p <.05\). (D) Pearson's r between distinctiveness/reconstruction error and hit rates across different presentation times. Error bars represent 95\% confidence interval from 1000 bootstrapping iterations. }
  \label{fig:fig4}
\end{figure*}

In Fig. 5B, we can see that although memory accuracy for all images increased with longer encoding times, images with larger reconstruction error benefited more from the longer encoding times (as indicated by the steeper slopes). A three-way repeated measures ANOVA confirmed these qualitative observations: In addition to the main effects of reconstruction error, distinctiveness, and encoding duration, there was a significant interaction between reconstruction error and encoding duration. That is, images with large reconstruction error significantly benefited more from longer encoding duration, relative to images with small reconstruction error (see Table S1). 

To directly measure the effect of reconstruction error on memory 
, we compared regression slopes relating encoding times to hit rates between the images with large vs. small reconstruction error. We bootstrapped the difference of these regression slopes between reconstruction error levels ( \(\beta_{Large RE} - \beta_{Small RE}\)) separately for each distinctiveness level and sequential pair of encoding times (Early: 34 ms vs. 84 ms; Late: 84 ms vs. 167 ms; see Methods).  
This analysis not only confirmed that images with larger reconstruction error indeed benefit more from longer encoding times, but also revealed a finer-grained temporal picture of this effect: As shown in Fig. 5C,  for images with large distinctiveness, the memory benefit of large reconstruction error was observed earlier (i.e., when encoding time was increased from 34 ms to 84 ms; mean \(\Delta\beta = 0.020,\ p = .045\)),  whereas for images with small distinctiveness, this reconstruction error benefit was only observed later (i.e., when encoding time was increased from 84 ms to 167 ms; mean \(\Delta\beta = 0.022,\ p = .001\)). 

Although Study 3 was conceived to reveal the differential effects of distinctiveness and reconstruction error in a categorical design (e.g., Large Distinctiveness and Small Reconstruction Error), next we turned to a finer-grained analysis to ask whether the ability of these models to predict behavior depends on time. We predict that if the memory benefit of reconstruction error is the result of deeper processing and thus requires more time to complete the computation, reconstruction error should account for more and more variance in memory as exposure time increases. To this end, for each duration, we correlated the behavioral hit rates at the level of individual images with their corresponding distinctiveness and reconstruction error. 
Indeed, as shown in Fig. 5D, we observed that the correlations between reconstruction error and hit rate increase monotonically as a function of exposure time (\(p < .001\) from 1000 bootstrapping iterations; see Methods) whereas this pattern is absent for distinctiveness (\(p = .526\) from 1000 bootstrapping iterations; see Methods).     

Together, these behavioral results further demonstrate that reconstruction error contributes to memorability by modulating encoding depth such that the memory benefit driven by reconstruction error only manifests itself when given enough encoding time. Moreover, distinctiveness- and reconstruction error-driven memorability effects exhibit differential temporal profiles and thus are likely to reflect functionally distinct processes in the mind and brain. 

\section*{Discussion}
We present a computational model that combines sparse coding with recent vision models based on deep convolutional neural networks and gives rise to a new quantitative measure -- i.e., compression-based reconstruction error -- of how perception modulates the strength of memory traces. 
In Studies 1 and 2, we show that reconstruction error predicts both memory accuracy and response times at the level of individual images. Critically, reconstruction error explains additional variance in both aspects of memory performance, beyond what can be explained by vision-only models (e.g., \cite{jaegle2019population,lin2019image}) -- including capturing all of the model-explained variance in response times. To further demonstrate that the modulation of memorability due to reconstruction error reflects a separate process relative to vision-only models,  we run an RSVP experiment and show that the effects of distinctiveness and reconstruction error on memorability have distinct temporal profiles: Memory performance increases to a more substantial degree for images that are harder to reconstruct (controlling for distinctiveness).
Finally, as exposure time increases, allowing a deeper analysis of the relevant scene features when necessary, reconstruction error becomes increasingly predictive of memory whereas the effect of distinctiveness stays more or less constant across durations. Together, these results not only establish compression-based reconstruction error as a previously unrecognized driver of memorability, but also suggest a perception to memory interface in which such reconstruction error modulates the encoding depth of incoming visual inputs. 

\subsection*{A signal for modulating depth of encoding}
By formalizing the interface between perception and memory with the sparse coding model, our work puts the level-of-processing theory in a new light. In the original and revised versions of the theory (\cite{craik1972levels,craik2002levels,craik2020remembering}), encoding depth has always been conceptualized as a continuum. Yet to date all empirical demonstrations of the effect of encoding depth on memory remain qualitative, through manipulating encoding depth with different orienting tasks (e.g., paying attention to the physical vs. semantic property of the stimuli; for a review, see \cite{cermak2014levels}). By demonstrating that reconstruction error, at the level of images, predicts memory accuracy, response latency during retrieval and the need for more encoding time, we suggest that reconstruction error can serve as an indication of the resource-rational depth of encoding, filling in a longstanding explanatory gap in the theory.

Multiple mechanisms are possible for how how reconstruction error can affect perceptual processing, leading to deeper or shallower analysis, and thus memory strength.
One possibility is suggested by the predictive coding framework \cite{friston2009predictive}, one version of which suggests the existence of predictive autoencoders that could compute the compressibility and reconstruction error of incoming information (e.g., \cite{rosenbaum2022relationship}). 
Another possibility is related to the idea of ‘analysis-by-synthesis’, which posits the reconstruction of sensory activations via a process of `synthesis' to be part of perceptual processing \cite{barrow1978recovering,olshausen201427,yuille2006vision,mumford1994neuronal}.
The current study suggests that such a synthesis process may also provide a learning signal for memory. 
Our results also join recent modeling work (e.g., \cite{benna2021place, dubreuil2014memory,wu2019dendrites}) to suggest the computational motif of sparse coding as a useful mechanism to formalize memory-related computations in the brain.
Future work can build on our framework to explore models that integrate compression and different forms of iterative processing guided by reconstruction error, and evaluate them as accounts of how percepts are cast into memory traces. 

\subsection*{A candidate for exploring the neurocomputational basis of the subsequent memory effect}
Decades of work in neuroimaging have clearly demonstrated that neural processing during perceptual encoding has a profound impact on what will later be remembered (e.g., the subsequent memory effect: \cite{brewer1998making, paller2002observing, wagner1998building}). Compared to items later forgotten, remembered items elicit greater activation in higher-level visual regions, frontoparietal attention regions and the medial temporal lobe 
\cite{kim2011neural} and higher pattern similarity across repetitions (\cite{xue2010greater,ward2013repetition,xue2018neural}). However, the computational processing that underlies these neural signals measured at the interface between perception and memory has been elusive. The empirical success of our SPC model in explaining multiple aspects of memory behavior suggests reconstruction error a plausible quantitative covariate to explain aspects of the neural signals underlying the subsequent memory effect. Future work should also explore how reconstruction error may selectively covary with activations from certain brain regions (e.g., frontoparietal regions and medial temporal lobe) vs. others (e.g., visual regions) to help elicit computational processes supported by these brain regions. \\

Beyond its role in driving memorability, reconstruction error can provide a more general priority signal that can be useful across multiple domains of cognition. These include the broader range of memory-related processes such as retrieval \cite{schacter2012adaptive,hemmer2009bayesian}, modulation of learning during development \cite{stahl2015observing}, and guiding the deployment of visual attention beyond bottom-up salience. Finally, our study also showcases the utility of `model-driven psychophysics'. Computational modeling allows us to generate quantitative predictions about behavioral performance and in turn, behavioral phenomena can help constrain and arbitrate between different models.

\section*{Methods}
\label{sec:methods}
\subsection*{Studies 1 and 2}

\subsubsection*{Dataset}
The dataset we used is from \cite{isola2014makes}. The dataset includes 2222 target images (with memorability scores measured from human observers; memorability score = hit rate - false alarm rate) and 8042 filler images (without memorability scores). For the reaction time (RT) measure, we only included RTs during the correctly recognized (i.e., hit) trials. After excluding outliers (+/- 2.5 standard deviations from the mean RT of all the correct trials), we averaged the remaining RTs for each target image to obtain one single measure of RT for the target image. The dataset covers a wide range of outdoor and indoor scenes. After removing duplicate images (i.e., images with different indices but identical content), we are left with 2221 target images and 8038 filler images. 

\subsubsection*{Quantifying distinctiveness}
We used the VGG-16 network trained for classifying both objects and scenes \cite{zhou2017places}. The model is trained to make 1365-way classification consisting of both object and scene categories (by merging the 1000 classes from ImageNet and the 365 classes from Places365-Standard). We used the Keras\cite{chollet2015keras} implementation of this model (https://github.com/GKalliatakis/Keras-VGG16-places365). We then sampled 7 layers across the hierarchy of the network to capture visual features at each layer. The 7 layers were maxpooling 1-5 and fully connected(fc) 1-2. For simplicity, we referred to these layers as Layer 1-7 in the manuscript. 

To quantify distinctiveness, we first passed all the images (resized from 256 \(\times\) 256 to 227 \(\times\) 227) in the Isola dataset (both targets and fillers) through the VGG-16 and extracted activation patterns across the 7 layers. Following our previous work \cite{lin2019image}, for each layer, we calculated distinctiveness as the Euclidean distance between each target image and its nearest neighbor (among all target and filler images) with respect to the feature space defined by the given layer. We also considered an alternative measure for quantifying DCNN network activations that was shown to be predictive of memorability scores (L2-Norm; \cite{jaegle2019population}). We found that both approaches yielded highly similar results (see Fig. S1). So we report results based on the nearest neighbor approach in the manuscript. 

\subsubsection*{Quantifying reconstruction error}
\emph{Sparse Coding}. A separate SPC model was trained for each of the 7 layers of the VGG-16 network \cite{zhou2017places}. SPC involves using linear combinations of a limited number of codewords (i.e., a codebook) to reconstruct inputs. 
To do so, we used LCA (locally competitive algorithm) \cite{rozell2008sparse} to implement sparse coding, which is a state-of-the-art method but computationally intensive algorithm. Main source of computational cost arises from the dimensionality of the inputs to be reconstructed. Thus, for the sake of computational tractability, we randomly sample $d=1000$ columns from the flattened feature activations for each layer. The training objective is then to reconstruct the resulting $1000-$dimensional feature vectors. In the language of SPC, this means that the dimension of each codeword is 1000.\footnote{We ensured that at this codeword length, the resulting reconstruction error measures were similar regardless of which 1000 units were randomly sampled and used to train SPC.}

Training SPC also requires a regularizer $\lambda$ for controlling how many codewords can be used to reconstruct a given feature vector. We choose $\lambda = 0.001$ based on a simple grid search to minimize reconstruction error. Because this regularization term is sensitive to the scale of feature vectors, and because we wished to use the same $\lambda$ for all layers, we applied a pre-processing step where we scaled feature vectors to be reconstructed using a layer-specific constant. This pre-processing step ensured that the same value of $\lambda$ worked equally well for all layers. \footnote{Note that this pre-processing only affects absolute magnitude of reconstruction errors within a layer, but does not affect relative order among the images with respect to that layer. For our study, the relevant metric is these relative differences of reconstruction errors, instead of the otherwise arbitrary absolute reconstruction error magnitudes.} 
Finally, we choose the codebook size (i.e., the number of codewords available to use for reconstructing a given input) to be $n = 500$. This codebook size is a reasonable trade-off between the computational cost of a large codebook in LCA and the reconstruction error incurred: setting smaller $n$s significantly increases reconstruction error. 

As mentioned above, a separate SPC model was trained for each layer of the VGG-16 network using all images in the Isola dataset, including both target images and filler images \footnote{We also tried training only on the filler images and the resulting reconstruction errors for the target images were largely the same (average \(r = 0.94\)).}. We used 800 iterations to train each model, where in each iteration a random batch of 50 images are used as training inputs. In each iteration, model weights are updated at most 500 times in an inner-loop, or less if the weights converged for the batch. To quantify reconstruction error, we calculated the Euclidean distance between the reconstructed 1000-dimensional vectors and the input 1000-dimensional activation vectors for each image. 

\subsubsection*{Evaluating relative contribution of distinctiveness and reconstruction error to memorability and response times}
We performed partial regression to evaluate the relative contributions of distinctiveness and reconstruction error to memorability/response times. More specifically, we first perform a simple linear regression (with intercept) that regresses the memorability scores/response times on distinctiveness and obtain residual $r_1$. Then, we perform another simple linear regression (with intercept) that regresses the reconstruction error on distinctiveness and obtain residual $r_2$. Finally, we correlate $r_1$ and $r_2$, and report statistics from this correlation. We also performed the same set of analyses to control for reconstruction error.  

\subsection*{Study 3}
\subsubsection*{Participants}
We recruited 65 participants via Prolific (www.prolific.co). As specified in our pre-registration (https://aspredicted.org/MFM\textunderscore R22), we excluded participants who did not complete all 12 blocks of the experiment (N = 19) or who did complete all 12 blocks but did not give a response on over 10\% of the trials on any given block (N = 1). Our final sample included 45 participants. The memory performance of all 45 participants was above chance (all d' \textgreater 0). 

\subsubsection*{Stimuli}
To investigate the roles distinctiveness and reconstruction error play in determining memorability, we sampled four groups of 48 target images each, with divergent profiles of distinctiveness and reconstruction error: 1) images with large distinctiveness and large reconstruction error, 2) images with large distinctiveness and small reconstruction error, 3) images with small distinctiveness and large reconstruction error, and 4) images with small distinctiveness and small reconstruction error. 

To sample these images, we started with using layer 5 nearest-1 neighbor distinctiveness as our distinctiveness measure because this yielded the highest correlation with the memorability scores (see Fig. 2A right panel). Then based on the partial regression results, we selected layer 7 reconstruction error from the sparse coding model as the reconstruction measure because it explained largest proportion of variance in memorability score beyond layer 5 distinctiveness (see Fig. 2B). In other words, we bench-marked distinctiveness and reconstruction error using the `best' performing combination. Using these benchmarks, we designated images in the top and bottom 30 percentile out of all the images (including targets and fillers) as `large' and `small' in terms of each measure (distinctiveness/reconstruction error). For each of the four image groups, we then sampled the 48 target images with the most extreme values considering both measures (e.g., for the large distinctiveness and large reconstruction error group, we sampled the 48 images with the largest sum of both percentiles; for the large distinctiveness and small reconstruction error group,  we sampled the 48 images with the largest difference between the distinctiveness percentile and the reconstruction error percentile). 

In addition to the 192 target images sampled in the way described above, we also randomly sampled 2304 images from the fillers in the Isola dataset to use as fillers in our experiment.

\subsubsection*{Design}
We adopted a within-participant 2 distinctiveness levels (High vs. Low) $\times$ 2 reconstruction error levels (High vs. Low) $\times$ 3 encoding durations (34, 84 or 167 ms) design. Each condition was presented as a separate block (i.e., 12 blocks in total, randomized across participants). During each block, half of the trials were target-present trials (i.e., the test image was presented in the RSVP stream) whereas the other half were target-absent trials (i.e., the test image was not presented in the RSVP stream). Trial order was randomized. Each block consists of 32 trials.

\subsubsection*{Procedure}
 Participants first completed a practice block of 20 trials with images not used in the actual experiment and all the practice images were presented for 167 ms each. At the end of each practice trial, they received feedback on whether their response was correct or not. After the practice block, participants were instructed to press the space bar to begin each experimental block and no feedback was given during the experimental blocks. Given the importance of timing for our experiment, images were pre-loaded at the beginning of each block to minimize the processing times during trial presentation. 

On each trial (following the trial structure of \cite{broers2018enhanced}, see Fig. 3B), participants first saw a 500-ms fixation cross, followed by a 200-ms blank screen. Then they saw a stream of 6 scene images presented back to back. Each image was shown for 34, 84 or 167 ms, depending on the block. If this was a target-present trial, the target image can be presented as the 2nd, 3rd, 4th, or 5th image in the RSVP stream (fully counter-balanced). After the RSVP stream, there was another 200-ms blank screen, followed by the test image, which was presented for 400 ms. Participants had a maximum of 10 seconds to respond either Yes (Press `1') or N (Press `0'). After the participant made a response, the next trial would start.  

\subsubsection*{Analysis}
The primary measure of interest is hit rate (percentage of correctly recognized target present trials), calculated separately for each of the 12 within-participant conditions. To assess the effect of encoding times as a function of reconstruction accuracy, we  conducted a three-way repeated-measures ANOVA with the dependent variable being hit rate and the three factors being distinctiveness, reconstruction error (RE) and encoding durations (Time). 

To further investigate how the interaction between distinctiveness, reconstruction error and encoding times affects memory, we ran bootstrapping analysis to resample the participants with replacement and only analyze trials where the encoding times are 34 ms and 84 ms (Early) or 84 ms and 167 ms (Late). For each participant, we first separated the trials into large and small distinctiveness. For each distinctiveness level, we fit two different linear regression models where Time was included as the predictor and hit rate as the dependent variable: one for images with large RE and the other for those with small RE. We then subtracted the fitted beta for Time for images with small RE from that for images with large RE. We next averaged this beta difference for all the participants in each bootstrapped sample. If our hypothesis holds, the slope for the linear regression model for large RE should be steeper than that for small RE and therefore the mean beta difference should be larger than 0. The bootstrapping procedure was repeated 1000 times and p value was calculated as the number of iterations where the mean beta difference is small than 0 (i.e., went against our hypothesis). 

The pre-registered analysis described above treated distinctiveness and reconstruction error as categorical variables and aggregated images into separate categories. In the following analysis, we aimed to test if the time-dependent effect of reconstruction error on memory also would hold when we evaluated the contributions of distinctiveness and reconstruction error to memorability of images presented at different exposure times in a continuous manner across all the target images. For each duration, we calculated the correlation between distinctiveness/reconstruction error and hit rate across all target images, respectively. To test statistical significance, we again ran bootstrapping analysis to resample the participants with replacement. For each iteration, we calculated correlations between distinctiveness/reconstruction error with hit rate across the three durations. We then fit two separate linear regression models, one for distinctiveness and the other for reconstruction error, to relate duration to correlation values. If our hypothesis holds, we should expect the the beta for the reconstruction error model is positive whereas the beta for the distinctiveness model should not be consistently positive. The bootstrapping procedure was repeated 1000 times and p value was calculated as the number of iterations where the beta is small than 0 (i.e., went against our hypothesis), separately for the reconstruction error model and the distinctiveness model.

\subsubsection*{Code and Data Availability}
Codes and data have been deposited on Github \\ (https://github.com/CNCLgithub/ReconMem).


 \bibliographystyle{elsarticle-num} 
 \bibliography{cas-refs}

\begin{thebibliography}{10}
\expandafter\ifx\csname url\endcsname\relax
  \def\url#1{\texttt{#1}}\fi
\expandafter\ifx\csname urlprefix\endcsname\relax\def\urlprefix{URL }\fi
\expandafter\ifx\csname href\endcsname\relax
  \def\href#1#2{#2} \def\path#1{#1}\fi

\bibitem{wagner1998building}
A.~D. Wagner, D.~L. Schacter, M.~Rotte, W.~Koutstaal, A.~Maril, A.~M. Dale,
  B.~R. Rosen, R.~L. Buckner, Building memories: remembering and forgetting of
  verbal experiences as predicted by brain activity, Science 281~(5380) (1998)
  1188--1191.

\bibitem{xue2018neural}
G.~Xue, The neural representations underlying human episodic memory, Trends in
  Cognitive Sciences 22~(6) (2018) 544--561.

\bibitem{craik1972levels}
F.~I. Craik, R.~S. Lockhart, Levels of processing: A framework for memory
  research, Journal of verbal learning and verbal behavior 11~(6) (1972)
  671--684.

\bibitem{schurgin2020psychophysical}
M.~W. Schurgin, J.~T. Wixted, T.~F. Brady, Psychophysical scaling reveals a
  unified theory of visual memory strength, Nature human behaviour 4~(11)
  (2020) 1156--1172.

\bibitem{chun2011memory}
M.~M. Chun, M.~K. Johnson, Memory: Enduring traces of perceptual and reflective
  attention, Neuron 72~(4) (2011) 520--535.

\bibitem{kurby2008segmentation}
C.~A. Kurby, J.~M. Zacks, Segmentation in the perception and memory of events,
  Trends in cognitive sciences 12~(2) (2008) 72--79.

\bibitem{favila2020transforming}
S.~E. Favila, H.~Lee, B.~A. Kuhl, Transforming the concept of memory
  reactivation, Trends in Neurosciences 43~(12) (2020) 939--950.

\bibitem{liu2021transformative}
J.~Liu, H.~Zhang, T.~Yu, L.~Ren, D.~Ni, Q.~Yang, B.~Lu, L.~Zhang, N.~Axmacher,
  G.~Xue, Transformative neural representations support long-term episodic
  memory, Science advances 7~(41) (2021) eabg9715.

\bibitem{libby2021rotational}
A.~Libby, T.~J. Buschman, Rotational dynamics reduce interference between
  sensory and memory representations, Nature neuroscience 24~(5) (2021)
  715--726.

\bibitem{serences2016neural}
J.~T. Serences, Neural mechanisms of information storage in visual short-term
  memory, Vision research 128 (2016) 53--67.

\bibitem{xu2017reevaluating}
Y.~Xu, Reevaluating the sensory account of visual working memory storage,
  Trends in Cognitive Sciences 21~(10) (2017) 794--815.

\bibitem{isola2014makes}
P.~Isola, J.~Xiao, D.~Parikh, A.~Torralba, A.~Oliva, What makes a photograph
  memorable?, IEEE Transactions on Pattern Analysis and Machine Intelligence
  7~(36) (2014) 1469--1482.

\bibitem{bainbridge2013intrinsic}
W.~A. Bainbridge, P.~Isola, A.~Oliva, The intrinsic memorability of face
  photographs., Journal of Experimental Psychology: General 142~(4) (2013)
  1323.

\bibitem{jaegle2019population}
A.~Jaegle, V.~Mehrpour, Y.~Mohsenzadeh, T.~Meyer, A.~Oliva, N.~Rust, Population
  response magnitude variation in inferotemporal cortex predicts image
  memorability, Elife 8 (2019) e47596.

\bibitem{khosla2015understanding}
A.~Khosla, A.~S. Raju, A.~Torralba, A.~Oliva, Understanding and predicting
  image memorability at a large scale, in: Proceedings of the IEEE
  international conference on computer vision, 2015, pp. 2390--2398.

\bibitem{lin2019image}
Q.~Lin, S.~R. Yousif, B.~Scholl, M.~M. Chun, Image memorability is driven by
  visual and conceptual distinctivenes, Journal of Vision 19~(10) (2019)
  290c--290c.

\bibitem{kramer2022features}
M.~A. Kramer, M.~N. Hebart, C.~I. Baker, W.~A. Bainbridge, The features
  underlying the memorability of objects, bioRxiv (2022).

\bibitem{baddeley1978trouble}
A.~D. Baddeley, The trouble with levels: A reexamination of craik and
  lockhart's framework for memory research. (1978).

\bibitem{treisman2014psychological}
A.~Treisman, The psychological reality of levels of processing, Levels of
  processing in human memory (2014) 301--330.

\bibitem{craik2020remembering}
F.~I. Craik, Remembering: An activity of mind and brain, Annual Review of
  Psychology 71 (2020) 1--24.

\bibitem{cermak2014levels}
L.~S. Cermak, F.~I. Craik, Levels of processing in human memory (PLE: Memory),
  Psychology Press, 2014.

\bibitem{bainbridge2020resiliency}
W.~A. Bainbridge, The resiliency of image memorability: A predictor of memory
  separate from attention and priming, Neuropsychologia 141 (2020) 107408.

\bibitem{bates2020efficient}
C.~J. Bates, R.~A. Jacobs, Efficient data compression in perception and
  perceptual memory., Psychological review 127~(5) (2020) 891.

\bibitem{schacter2012adaptive}
D.~L. Schacter, Adaptive constructive processes and the future of memory.,
  American Psychologist 67~(8) (2012) 603.

\bibitem{hemmer2009bayesian}
P.~Hemmer, M.~Steyvers, A bayesian account of reconstructive memory, Topics in
  Cognitive Science 1~(1) (2009) 189--202.

\bibitem{olshausen1996emergence}
B.~A. Olshausen, D.~J. Field, Emergence of simple-cell receptive field
  properties by learning a sparse code for natural images, Nature 381~(6583)
  (1996) 607--609.

\bibitem{olshausen1997sparse}
B.~A. Olshausen, D.~J. Field, Sparse coding with an overcomplete basis set: A
  strategy employed by v1?, Vision research 37~(23) (1997) 3311--3325.

\bibitem{benna2021place}
M.~K. Benna, S.~Fusi, Place cells may simply be memory cells: Memory
  compression leads to spatial tuning and history dependence, Proceedings of
  the National Academy of Sciences 118~(51) (2021) e2018422118.

\bibitem{lewicki2002efficient}
M.~S. Lewicki, Efficient coding of natural sounds, Nature neuroscience 5~(4)
  (2002) 356--363.

\bibitem{zemel1993developing}
R.~Zemel, G.~E. Hinton, Developing population codes by minimizing description
  length, Advances in neural information processing systems 6 (1993).

\bibitem{rozell2008sparse}
C.~J. Rozell, D.~H. Johnson, R.~G. Baraniuk, B.~A. Olshausen,
  \href{https://doi.org/10.1162/neco.2008.03-07-486}{Sparse coding via
  thresholding and local competition in neural circuits}, Neural Comput.
  20~(10) (2008) 2526–2563.
\newblock \href {https://doi.org/10.1162/neco.2008.03-07-486}
  {\path{doi:10.1162/neco.2008.03-07-486}}.
\newline\urlprefix\url{https://doi.org/10.1162/neco.2008.03-07-486}

\bibitem{rumelhart1985learning}
D.~E. Rumelhart, G.~E. Hinton, R.~J. Williams, Learning internal
  representations by error propagation, Tech. rep., California Univ San Diego
  La Jolla Inst for Cognitive Science (1985).

\bibitem{gregor2010learning}
K.~Gregor, Y.~LeCun, Learning fast approximations of sparse coding, in:
  Proceedings of the 27th international conference on international conference
  on machine learning, 2010, pp. 399--406.

\bibitem{zhou2017places}
B.~Zhou, A.~Lapedriza, A.~Khosla, A.~Oliva, A.~Torralba, Places: A 10 million
  image database for scene recognition, IEEE transactions on pattern analysis
  and machine intelligence 40~(6) (2017) 1452--1464.

\bibitem{DBLP:journals/corr/SimonyanZ14a}
K.~Simonyan, A.~Zisserman, \href{http://arxiv.org/abs/1409.1556}{Very deep
  convolutional networks for large-scale image recognition}, in: Y.~Bengio,
  Y.~LeCun (Eds.), 3rd International Conference on Learning Representations,
  {ICLR} 2015, San Diego, CA, USA, May 7-9, 2015, Conference Track Proceedings,
  2015.
\newline\urlprefix\url{http://arxiv.org/abs/1409.1556}

\bibitem{berger1971rate}
T.~Berger, Rate distortion theory: A mathematical basis for data compression,
  Prentice-Hall., 1971.

\bibitem{cover1991elements}
T.~M. Cover, J.~A. Thomas, Elements of Information Theory, John Wiley \& Sons,
  1991.

\bibitem{mackay2003information}
D.~J. MacKay, Information Theory, Inference, and Learning Algorithms, Cambridge
  University Press, 2003.

\bibitem{hinton1995wake}
G.~E. Hinton, P.~Dayan, B.~J. Frey, R.~M. Neal, The" wake-sleep" algorithm for
  unsupervised neural networks, Science 268~(5214) (1995) 1158--1161.

\bibitem{kahana1999response}
M.~Kahana, G.~Loftus, Response time versus accuracy in human memory, The nature
  of cognition (1999) 322--384.

\bibitem{bylinskii2015intrinsic}
Z.~Bylinskii, P.~Isola, C.~Bainbridge, A.~Torralba, A.~Oliva, Intrinsic and
  extrinsic effects on image memorability, Vision research 116 (2015) 165--178.

\bibitem{vincent1996relations}
A.~Vincent, F.~I. Craik, J.~J. Furedy, Relations among memory performance,
  mental workload and cardiovascular responses, International Journal of
  Psychophysiology 23~(3) (1996) 181--198.

\bibitem{ragland2003levels}
J.~D. Ragland, S.~T. Moelter, C.~McGrath, S.~K. Hill, R.~E. Gur, W.~B. Bilker,
  S.~J. Siegel, R.~C. Gur, Levels-of-processing effect on word recognition in
  schizophrenia, Biological psychiatry 54~(11) (2003) 1154--1161.

\bibitem{broers2018enhanced}
N.~Broers, M.~C. Potter, M.~R. Nieuwenstein, Enhanced recognition of memorable
  pictures in ultra-fast rsvp, Psychonomic bulletin \& review 25~(3) (2018)
  1080--1086.

\bibitem{craik2002levels}
F.~I. Craik, Levels of processing: Past, present... and future?, Memory
  10~(5-6) (2002) 305--318.

\bibitem{friston2009predictive}
K.~Friston, S.~Kiebel, Predictive coding under the free-energy principle,
  Philosophical transactions of the Royal Society B: Biological sciences
  364~(1521) (2009) 1211--1221.

\bibitem{rosenbaum2022relationship}
R.~Rosenbaum, On the relationship between predictive coding and
  backpropagation, Plos one 17~(3) (2022) e0266102.

\bibitem{barrow1978recovering}
H.~Barrow, J.~Tenenbaum, Recovering intrinsic scene characteristics, Comput.
  Vis. Syst 2~(3-26) (1978) 2.

\bibitem{olshausen201427}
B.~A. Olshausen, G.~Mangun, M.~Gazzaniga, Perception as an inference problem,
  MIT press Cambridge, MA, 2014.

\bibitem{yuille2006vision}
A.~Yuille, D.~Kersten, Vision as bayesian inference: analysis by synthesis?,
  Trends in cognitive sciences 10~(7) (2006) 301--308.

\bibitem{mumford1994neuronal}
D.~Mumford, Neuronal architectures for pattern-theoretic problems, Large-scale
  neuronal theories of the brain (1994) 125--152.

\bibitem{dubreuil2014memory}
A.~M. Dubreuil, Y.~Amit, N.~Brunel, Memory capacity of networks with stochastic
  binary synapses, PLoS computational biology 10~(8) (2014) e1003727.

\bibitem{wu2019dendrites}
X.~Wu, G.~C. Mel, D.~Strouse, B.~W. Mel, How dendrites affect online
  recognition memory, PLoS computational biology 15~(5) (2019) e1006892.

\bibitem{brewer1998making}
J.~B. Brewer, Z.~Zhao, J.~E. Desmond, G.~H. Glover, J.~D. Gabrieli, Making
  memories: brain activity that predicts how well visual experience will be
  remembered, Science 281~(5380) (1998) 1185--1187.

\bibitem{paller2002observing}
K.~A. Paller, A.~D. Wagner, Observing the transformation of experience into
  memory, Trends in cognitive sciences 6~(2) (2002) 93--102.

\bibitem{kim2011neural}
H.~Kim, Neural activity that predicts subsequent memory and forgetting: a
  meta-analysis of 74 fmri studies, Neuroimage 54~(3) (2011) 2446--2461.

\bibitem{xue2010greater}
G.~Xue, Q.~Dong, C.~Chen, Z.~Lu, J.~A. Mumford, R.~A. Poldrack, Greater neural
  pattern similarity across repetitions is associated with better memory,
  Science 330~(6000) (2010) 97--101.

\bibitem{ward2013repetition}
E.~J. Ward, M.~M. Chun, B.~A. Kuhl, Repetition suppression and multi-voxel
  pattern similarity differentially track implicit and explicit visual memory,
  Journal of Neuroscience 33~(37) (2013) 14749--14757.

\bibitem{stahl2015observing}
A.~E. Stahl, L.~Feigenson, Observing the unexpected enhances infants’
  learning and exploration, Science 348~(6230) (2015) 91--94.

\bibitem{chollet2015keras}
F.~Chollet, et~al., Keras, \url{https://keras.io} (2015).

\end{thebibliography}

\newpage
\renewcommand{\thetable}{S\arabic{table}}

\begin{table}[hbt!]
\centering
 \begin{tabular}{||c c c c c c||} 
 \hline
 Effect & DFn & DFd & F & p & \(\eta^{2}\) \\ [0.5ex] 
 \hline\hline
Distinctiveness (Dist.) & \(1\) & \(44\) & \(197.187\)& \(< .001\)  & \(0.353\)\\ 
 Reconstruction error (RE) & \(1\) & \(44\) & \(27.408\)& \(< .001\) & \(0.029\)\\
 Time & \(2\) & \(88\) & \(106.498\) & \(< .001\)& \(0.285\)\\
 Dist. \(\times \) RE & \(1\) & \(44\) &\(0.160\) & &  \(0.000\)\\
 Dist. \(\times \) Time & \(2\)& \(88\) & \(4.414\) &  \(.015\) &  \(0.001\)\\
 RE \(\times \) Time &\(2\) & \(88\) & \(4.192\) & \(.018\) & \(0.001\)\\
 Dist. \(\times \)RE \(\times \) Time & \(2\) & \(88\) & \(3.909\) & \(.024\) &\(0.001\)\\ [1ex] 
 \hline
 \end{tabular}
 \caption{\label{tab:table1}To formally test how distinctiveness and reconstruction error influence the effect of encoding time on memory performance, we ran a three-way repeated-measures ANOVA with the dependent variable being hit rate and the three factors being distinctiveness (Dist), reconstruction error (RE) and encoding duration (Time). The table shows the results. As expected, there were main effects of distinctiveness, reconstruction error and encoding duration. More critically, we also saw an interaction between reconstruction error and encoding duration, meaning that images with large reconstruction error benefited even more from longer encoding times, relative to images with smaller reconstruction error. Interestingly, we also observed an interaction between distinctiveness and encoding duration and a three-way interaction between distinctiveness, reconstruction error and encoding duration. These interactions can arise from the differential time-courses of the memory benefit (see Fig. 5C) brought by large reconstruction error as a function of the distinctiveness levels.}
\end{table}

\newpage
\renewcommand{\thefigure}{S\arabic{figure}}
\setcounter{figure}{0}
\begin{figure}[hbt!]
  \centering
  \includegraphics[width=1\textwidth]{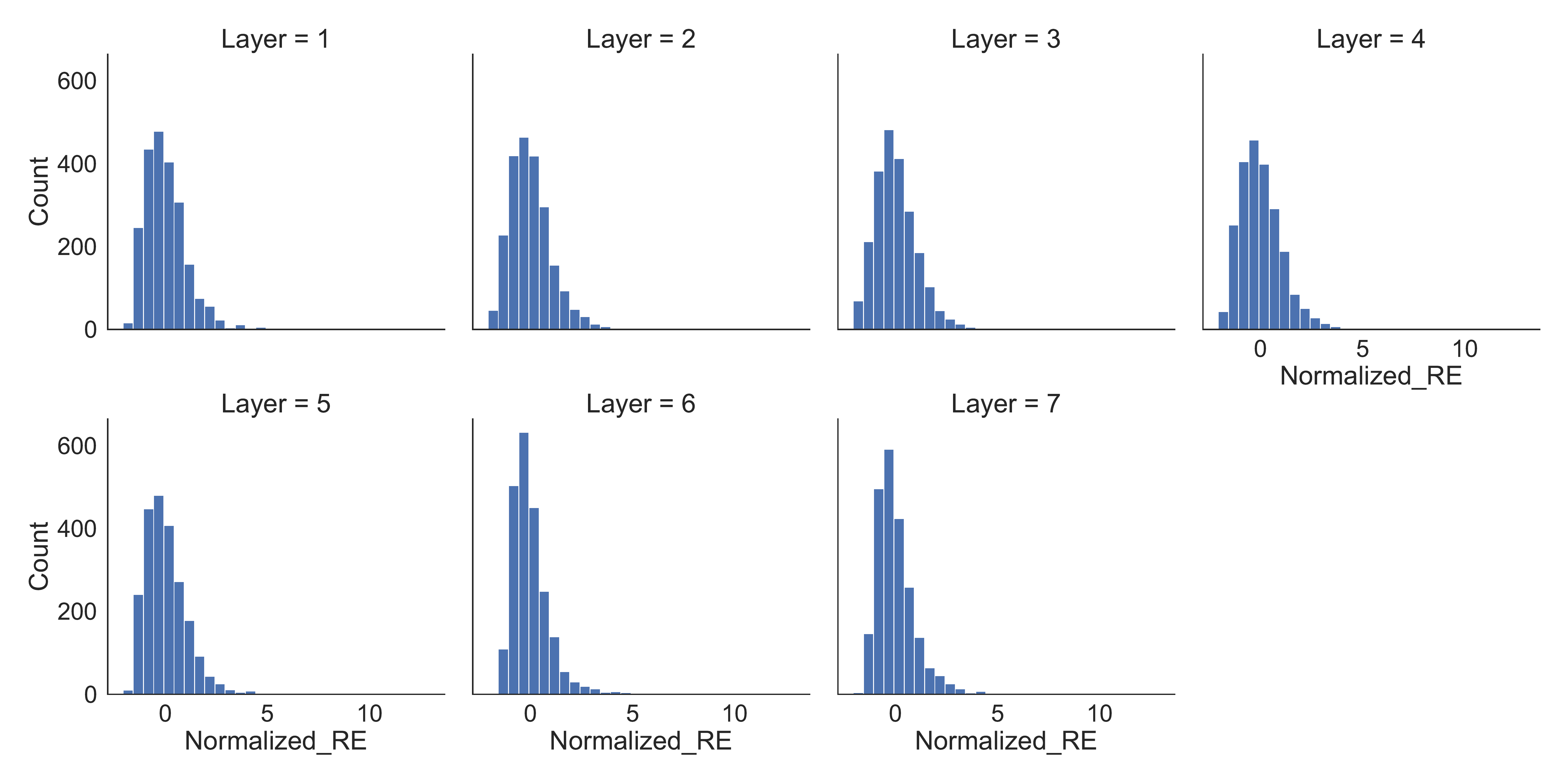}
  \caption{The distributions of normalized reconstruction errors for the 2221 target images from the seven sparse coding models (each trained to reconstruct a different layer of VGG-16).}
  \label{fig:suppfig1}
\end{figure}

\begin{figure}[hbt!]
  \centering
  \includegraphics[width=1\textwidth]{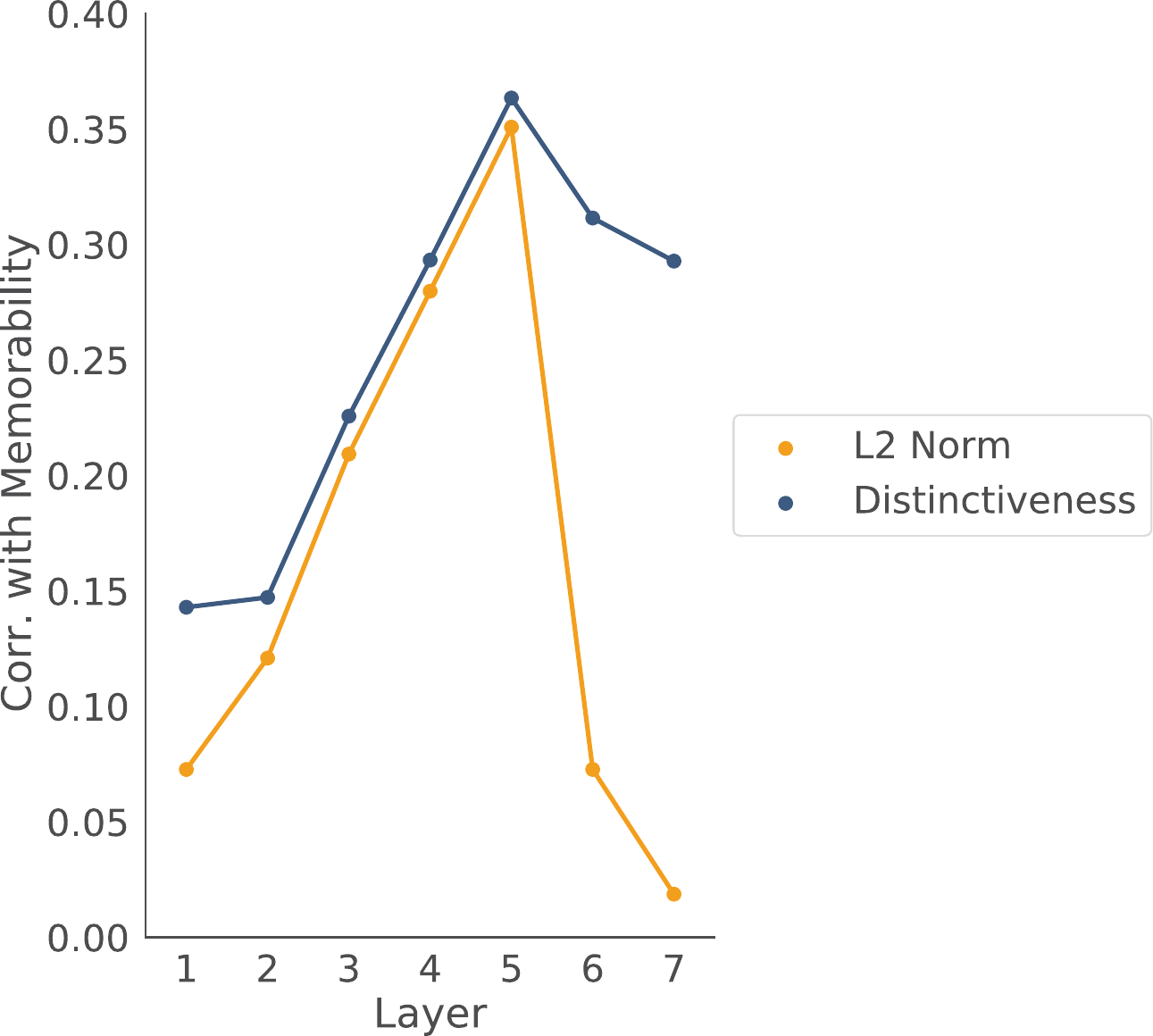}
  \caption{Correlations between memorability and two different measures derived from activation in VGG-16: Distinctiveness (euclidean distance to nearest neighbor) and L2 Norm (L2 norm of the activation vector).}
  \label{fig:suppfig2}
\end{figure}

\begin{figure}[hbt!]
  \centering
  \includegraphics[width=1\textwidth]{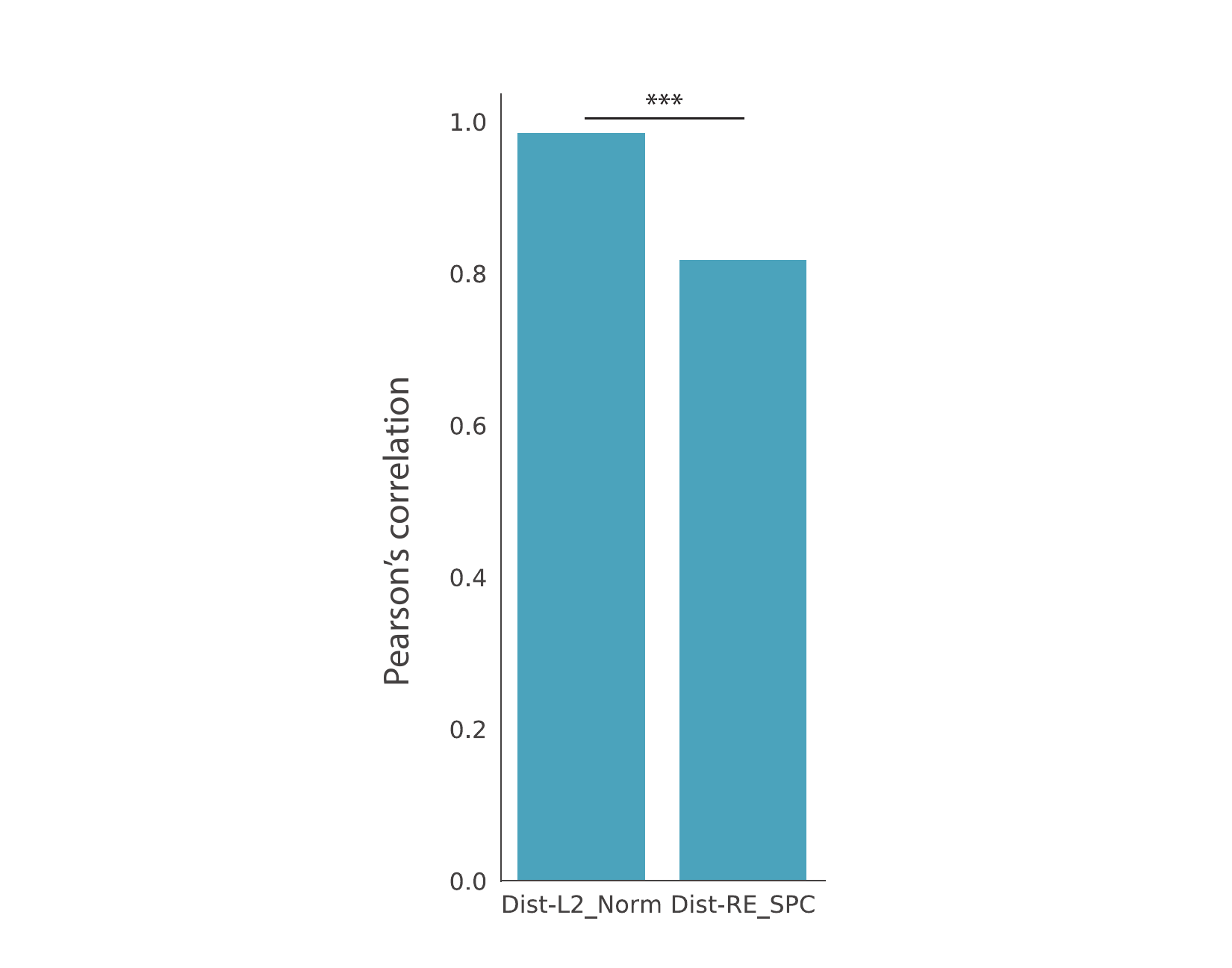}
  \caption{Pearson's correlations between different measures for Layer 5. Dist(Distinctiveness): euclidean distance to nearest neighbor. L2\textunderscore  Norm: L2 norm of the activation vector. RE\textunderscore SPC: reconstruction error from the sparse coding model. Statistical significance was assessed with Williams' t test (for two correlation values sharing one common variable on the same population). ***: p \(< .001\) }
  \label{fig:suppfig3}
\end{figure}





\end{document}